\documentclass[pdftex,twocolumn,epjc3]{svjour3}          

\usepackage[T1]{fontenc}
\usepackage[english]{babel}
 \addto\captionsenglish{
                        \renewcommand{\figurename} {FIG.}%
                        \renewcommand{\tablename}  {TABLE}%
                        \renewcommand{\appendixname}{APPENDIX}%
                       }
\usepackage[numbers,sort&compress]{natbib} 

\usepackage{graphicx}
 \graphicspath{{graphics/}}
\usepackage{multirow}

\usepackage[caption=false]{subfig}
\usepackage[export]{adjustbox}

\usepackage{amsmath}
\usepackage{amssymb}
\usepackage[version=3]{mhchem}
\usepackage{mathrsfs}
\usepackage{bm}
\usepackage{todonotes}
\usepackage{upgreek}
\usepackage{soul}
\usepackage{doi} 
\hypersetup{
            colorlinks,
            linkcolor={red!70!black}  ,
            citecolor={green!70!black},
            urlcolor ={blue!70!black}
           }

\usepackage{comment} 
\usepackage[switch]{lineno}
\usepackage{xspace}

\begin{document}


 \journalname{Eur. Phys. J. C}
 \title{First measurement of Gallium Arsenide as a low-temperature calorimeter}

 \author{
              D.~L.~Helis~\thanksref{INFNlngs,corrAuth}
         \and A.~Melchiorre~\thanksref{INFNlngs,UnivAQ,corrAuth}
         \and A.~Puiu~\thanksref{INFNlngs,corrAuth}
         \and G.~Benato~\thanksref{GSSI, INFNlngs}
         \and P.~Carniti~\thanksref{Mib}
         \and A.~Continenza~\thanksref{UnivAQ}
         \and N.~Di Marco\thanksref{GSSI,INFNlngs}
         \and A.~Ferella~\thanksref{UnivAQ,INFNlngs}
         \and C.~Ferrari~\thanksref{GSSI,INFNlngs}
         \and F.~Giannesi~\thanksref{UnivAQ}
         \and C.~Gotti~\thanksref{Mib}
         \and E.~Monticone~\thanksref{INRIM}
         \and L.~Pagnanini~\thanksref{GSSI,INFNlngs}
         \and G.~Pessina~\thanksref{Mib}
         \and S.~Pirro~\thanksref{INFNlngs}
         \and G.~Profeta~\thanksref{UnivAQ}
         \and M.~Rajteri~\thanksref{INRIM}
         \and P.~Settembri~\thanksref{UnivAQ}
         \and A.~Shaikina~\thanksref{GSSI,INFNlngs}
         \and C.~Tresca~\thanksref{CNR}
         \and D.~Trotta~\thanksref{Mib}   }
    \institute{INFN -- Laboratori Nazionali del Gran Sasso, Assergi, I-67100 L'Aquila, Italy\label{INFNlngs}
            \and Dipartimento di Scienze Fisiche e Chimiche, Universit\`a degli Studi dell'Aquila, I-67100 L'Aquila, Italy\label{UnivAQ}
            \and Gran Sasso Science Institute, I-67100 L'Aquila, Italy\label{GSSI}
            \and INFN and Università degli studi di Milano-Bicocca, I-20126 Milano, Italy \label{Mib}
            \and Istituto Nazionale di Ricerca Metrologica Torino, I-10135 Torino, Italy\label{INRIM}
            \and CNR-SPIN, Università degli studi dell’Aquila, I-67100 L’Aquila, Italy\label{CNR}    \\
         }
           \newcommand{\emailaddress}{andrei.puiu@lngs.infn.it, \\ andrea.melchiorre@lngs.infn.it, dounia.helis@lngs.infn.it}
           \thankstext{corrAuth}{Corresponding Authors: \href{mailto:\emailaddress}{\emailaddress}}

 \twocolumn
 \maketitle

 \begin{abstract}

In this paper we present the first measurement of a Gallium Arsenide crystal working as low-temperature calorimeter for direct Dark Matter (DM) searches within the DAREDEVIL (DARk-mattEr DEVIces for Low energy detection) project. In the quest for direct dark matter detection, innovative approaches to lower the detection thershold and explore the sub-GeV mass DM range, have gained high relevance in the last decade. This study presents the pioneering use of Gallium Arsenide (GaAs) as a low-temperature calorimeter for probing the dark matter-electron interaction channel. Our experimental setup features a GaAs crystal at ultralow temperature of 15 mK, coupled with a Neutron Transmutation Doped (NTD) thermal sensor for precise energy estimation. This configuration is the first step towards the detection of single electrons scattered by dark matter particles within the GaAs crystal, to significantly improve the sensitivity to low-mass dark matter candidates.
This paper presents a comprehensive study of the detector's response, a calibration spectrum obtained using $\alpha$ events and X-ray events, which serves as a benchmark for evaluating the detector's performance, sensitivity, and threshold. These findings not only demonstrate the feasibility of using GaAs as cryogenic calorimeter,  but also open new avenues for investigating the elusive nature of dark matter through innovative direct detection techniques.

 \end{abstract}

 \keywords{Low-Temperature calorimeters, Dark Matter searches, Gallium Arsenide }

 \section{Introduction}

The pursuit to elucidate the enigmatic nature of dark matter, which accounts for approximately 27$\%$ of the universe's total mass-energy composition, has spurred innovative strategies in particle physics. One such pioneering approach is the direct detection of dark matter through its scattering off electrons in detector materials. This method stands at the forefront of efforts to probe dark matter's non-gravitational interactions, a quest that has predominantly focused on interactions between dark matter and nucleons.

Indeed, current worldwide experimental efforts in the field of direct dark matter search primarily focus on the Weakly Interacting Massive Particle (WIMP) paradigm within the GeV-TeV mass range, employing various approaches and techniques. The increasing sensitivity of these experiments has resulted in the exclusion of significant portions of the phase space. While future experiments on a multi-ton scale (such as LZ, XENONnT, DS, DARWIN) are expected to approach the so-called neutrino floor, there is a growing emphasis on the search for sub-GeV dark matter. Numerous models of light dark matter have been proposed, including asymmetric dark matter, freeze-in, strong dynamics, kinematic threshold, among others, which hypothetically expand the search window to include dark matter particles at the eV scale \cite{XENON:2020rca}.

The interaction of dark matter with electrons, known as dark matter-electron scattering, is particularly revealing as it opens a window to detect and analyze dark matter particles with much lower masses than those accessible through nucleon interactions \cite{Billard:2021uyg}, thereby broadening the scope of dark matter searches to include candidates in the sub-GeV range. 
Furthermore, understanding DM-electron scattering dynamics can provide insights into the potential mechanisms through which dark matter interacts with ordinary matter, offering clues to the scattering processes that mediate these interactions \cite{Billard:2021uyg}.\\

The search for light DM candidates in the last decade has provided an astonishing number of new results. The XENONnT collaboration's findings \cite{XENON:2022ltv} have placed tight constraints on the interaction cross-sections, reinforcing the potential of electronic recoil detection to explore new regions of the dark matter parameter space. The large-volume noble liquid detectors like LUX \cite{PhysRevLett.122.131301}, PANDA-X \cite{PandaX-II:2021nsg}, and Dark-Side-50 \cite{Franco:2023sjx} are also sensitive to sub-GeV DM.
Other dark matter experiments such as DAMIC-M \cite{DAMIC-M:2023gxo}, EDELWEISS \cite{EDELWEISS:2020fxc}, SENSEI \cite{SENSEI:2023zdf}, Super-CDMS \cite{SuperCDMS:2018mne} have set limits very stringent limits to DM-electron scattering reaching the MeV/$c^2$ mass region.\\

In this panorama, the DAREDEVIL\cite{DAREDEVIL_WHitePaper_2024} project aims to take the first step towards creating a low-threshold, highly-resolution device for future experiments targeting the search for light dark matter. We plan to develop a series of prototype detectors by coupling state-of-the-art cryogenic sensors to small- or zero-gap materials, such as Dirac, Weyl crystals and superconductors \cite{Derenzo:2016fse,Derenzo:2018plr,Vasiukov:2019hwn}. The last category is of particular interest for a specific material (GaAs) which is both a scintillator with O(eV) gap (smaller than the classical NaI or CsI), a
semiconductor and a polar crystal (having high sensitivity to dark photon absorption \cite{Knapen:2017ekk}). In this case a possible double read-out
could be implemented, reading both the light signal and the NTL-enhanced phonon signal. 

Theoretical models have suggested that such low-mass dark matter particles might scatter off electrons, producing detectable electronic recoils \cite{Billard:2021uyg}. This premise is supported by extensive theoretical work \cite{PhysRevD.85.076007}, which is crucial in demonstrating the potential of semiconductor detectors to reveal interactions involving dark matter particles with masses lower than the MeV scale. 
The TESSARACT experiment is also considering the use of this crystal for their research \cite{Essig:2022dfa}.

DAREDEVIL's innovative approach not only aligns with the latest theoretical and experimental advancements but also pioneers a novel strategy to probe the elusive MeV dark matter sector. This significant advancement in dark matter research promises to shed light on the properties and interactions of dark matter particles within this challenging mass range.\\

 \section{Dark Matter detection with GaAs}
Gallium Arsenide has emerged as a forefront candidate in the search for low-mass dark matter due to its unique combination of physical and electronic properties, particularly when operated as a calorimeter at milliKelvin temperatures. The intrinsic band gap of GaAs, approximately 1.42 eV at room temperature, is crucial for its sensitivity to low-energy excitations, enabling the detection of sub-MeV dark matter particles through electron recoil processes.
Although GaAs was predicted to have an higher DM-electron fiducial cross section with respect to lower-gap semiconductors and metals,  it outperforms all the other detectors based on insulators.
In addition, it still can offers other electronic detection channels which can be further exploited, like the Migdal effect\cite{PhysRevD.105.015014}.
The electronic sensitivity can be further enhanced by the ability to modify the band gap through doping, as detailed in \cite{PhysRevD.96.016026,Budnik:2017sbu}, where the introduction of specific dopants can tailor the electronic and optical properties of GaAs to optimize its response to dark matter interactions. Doping not only adjusts the conductivity and creates color centers, which are pivotal for the scintillation process, but also introduces additional energy levels within the band gap, enhancing the detection of even smaller energy deposits from dark matter interactions.
Indeed, recently it was proposed DM-dopant atom as a new DM detection strategy\cite{PhysRevD.109.055009} to explore
sub-MeV DM coupling to the SM via a light mediator, and kinetically mixed sub-eV dark-photon DM.
For lower energy deposition, below the band-gap, GaAs still offer interesting detection channels. In this energy region the energy loss is dominated by phonon excitations, and in particular optical phonons in polar materials, like GaAs. Indeed, considering DM-phonon scattering, including multiphonon processes, allow GaAs to reach lower DM masses.\cite{PhysRevD.105.015014}

In addition, thanks to its reduced heat capacity operating GaAs as a calorimeter at milliKelvin temperatures makes potentially possibile to detect energy depositions of the order of a few eV. 

Recent studies have moreover demonstrated that GaAs has scintillation properties and its yields and optical characteristics of GaAs can be precisely adjusted through dopant selection, enabling a dual-detection mechanism in conjunction with cryogenic light detectors \cite{Vasiukov:2019hwn,Griffin:2018bjn}. This approach, which effectively combines the detection of phonons generated from particle interactions with scintillation light, allows for event by event particle identification. Moreover, GaAs's versatility and the possibility of engineering its properties to meet specific detection requirements, from bandgap adjustment through doping to scintillation response optimization highlight its potential in pioneering the search for low mass dark matter \cite{Griffin:2018bjn,Knapen:2017ekk}.

In the following paragraphs, we will present the very first measurement of a low-temperature detector using GaAs as the absorber crystal coupled to a state-of-the-art NTD thermistor used to detect the phonon signal.

 \section{Experimental Setup}
For this first measurement of GaAs as a cryogenic calorimeter, we used a 2-inch diameter and 0.5 mm thick wafer (5.35 g). The wafer was equipped with a 3 $\times$ 0.6 $\times$ 0.4 mm Neutron Transmutation Doped (NTD) sensor using three epoxy glue spots onto the polished surface of the GaAs crystal. This sensor underwent a precise irradiation process to achieve a doping level corresponding to a T$_0$ parameter of about 4~K, ensuring enhanced sensitivity and stability for our measurements given its resistance dependence of temperature $R(T)=R_0 \cdot e^{-(T/T_0)^{1/2}}$ \cite{Haller1984}.
 \begin{figure}
    \centering
    \includegraphics[width=0.45\textwidth]{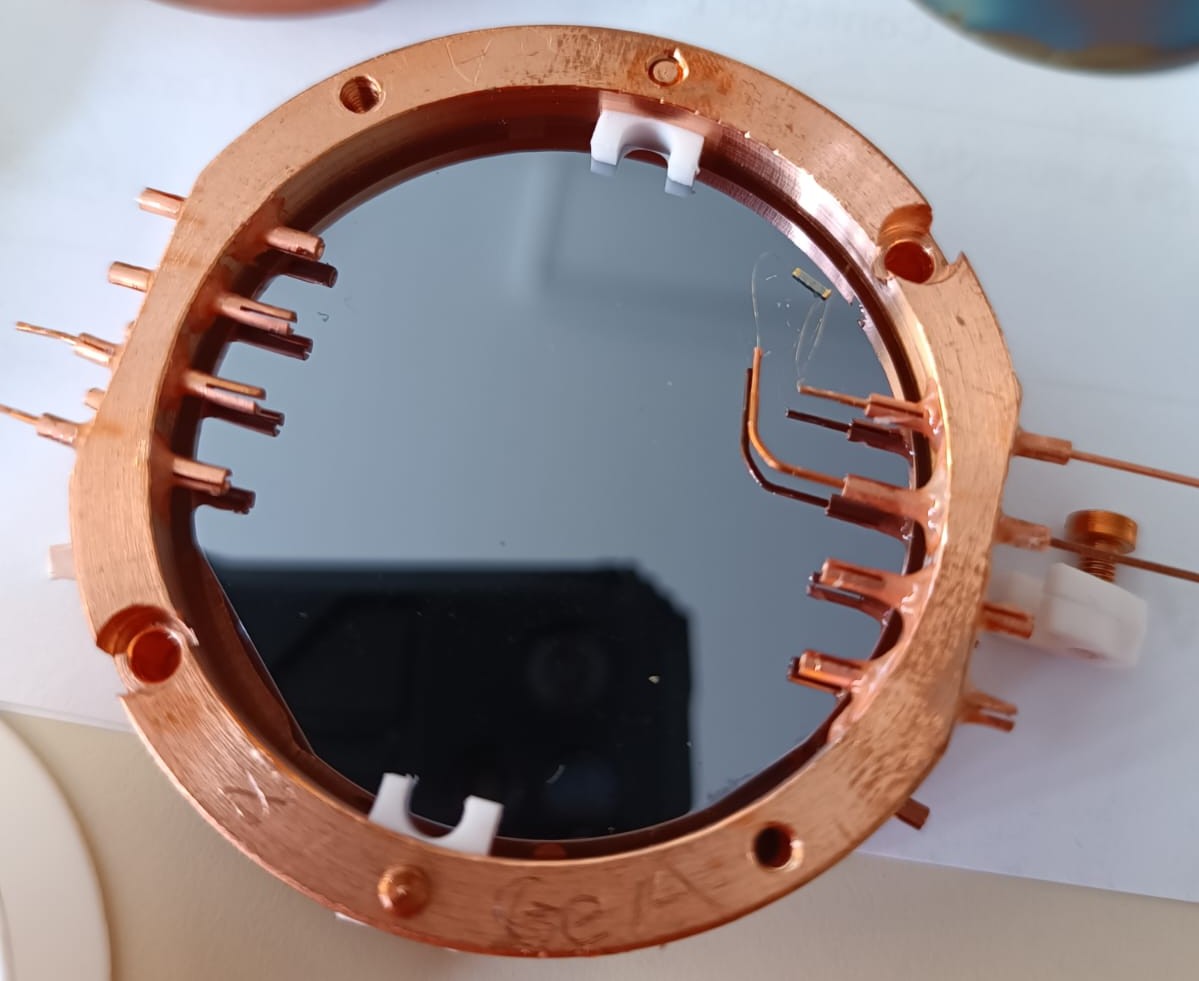}
    \caption{GaAs detector mounted with PTFE supports on the copper frame. On the right-hand side, the NTD is glued and connected via two 25 $\mu$m thick gold wires to read the signal. The whole frame was then coupled to the 10 mK stage of the dilution refrigerator.}
    \label{fig:set}
\end{figure}
The GaAs wafer was mounted on polytetrafluoroethylene (PTFE) supports within a copper frame. This configuration was chosen for its optimal structural integrity and thermal conductivity. The copper frame was directly coupled to the mixing chamber of a pulse tube (PT)-assisted dilution refrigerator IETI \cite{ieti}, located in Hall C at the Laboratori Nazionali del Gran Sasso, reaching a base temperature of 15 mK. To perform calibration, we employed a $^{238}$U source with a total activity of approximately 1 Bq to illuminate the GaAs wafer along with a $^{55}$Fe source to illuminate the detector with 5.9 keV X-rays from the Mn K-$\alpha$. This setup was specifically designed to expose the wafer to high energy 4.198 MeV $\alpha$ particles, alongside $\beta$ and $\gamma$ events with energies reaching up to 2 MeV and lower energy X-rays.

To complete the setup, the NTD sensor was connected with the help of golden wires to the refrigerator's wiring, enabling signal readout through electronics positioned at room temperature. The readout electronics of our system consist of low-noise DC-coupled front-end boards operated at room temperature and high-resolution digitizers  \cite{Arnaboldi_2018,Carniti:2022coa} . These boards are equipped with low-noise DC coupled front-end electronics to ensure optimal performance.

 \section{Data Analysis and Results}
We conducted a  12-hour long calibration using the \textsuperscript{238}U and \textsuperscript{55}Fe source where the data is acquired using a MATLAB program specifically developed to handle the continuous data stream. This program digitizes the data at a frequency of 10 kHz, with a cut-off frequency of 500 Hz to eliminate higher frequency noise. The digitized data is then stored on disk as binary files for further offline analysis.

\begin{figure}[h]
    \centering
    \includegraphics[width=0.5\textwidth]{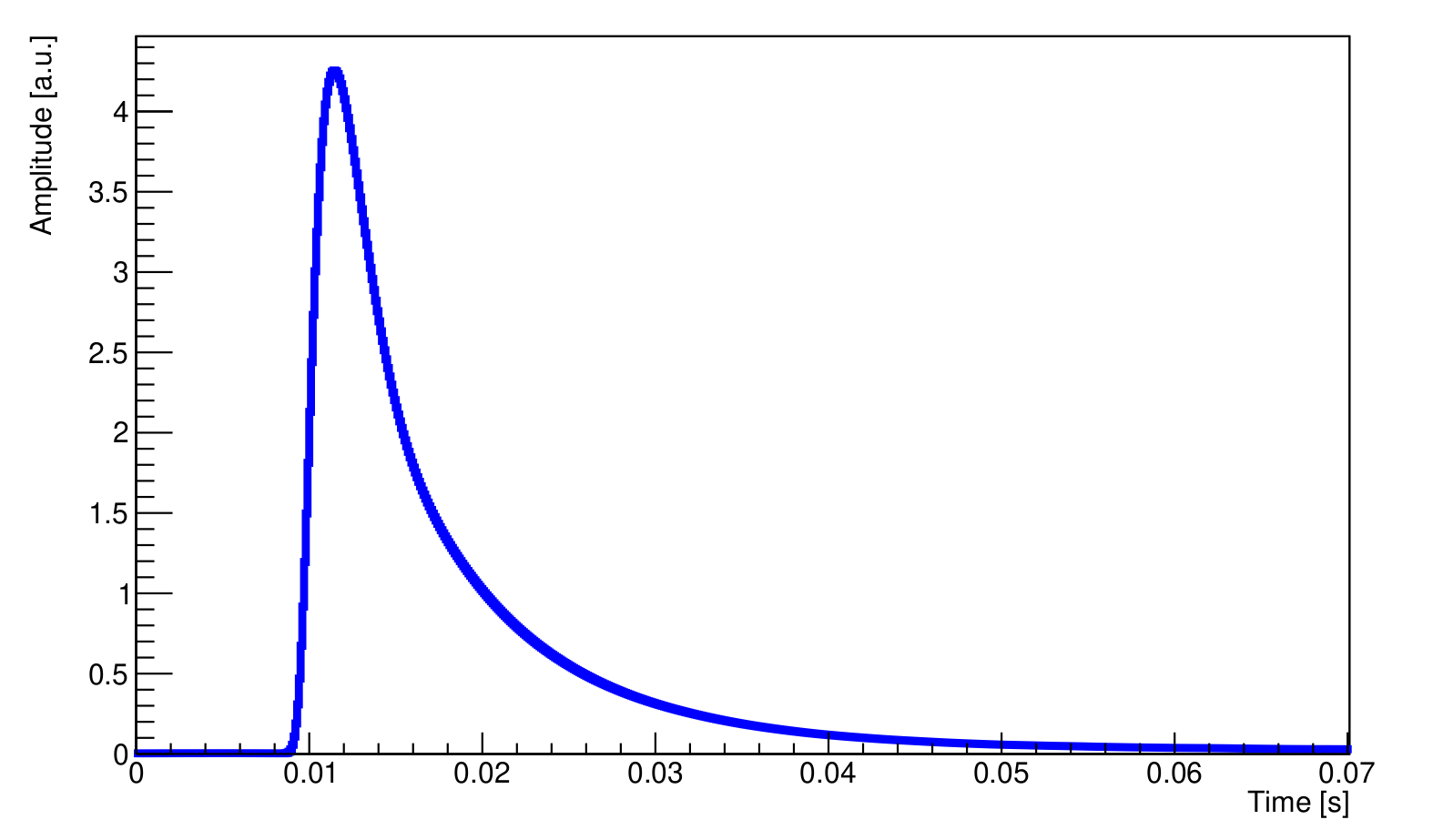}
    \caption{Average signal of the studied GaAs detector, featuring a rise time of 1.2 ms and a decay time of 10.8 ms. The entire signal is contained within a 70 ms window.}
    \label{fig:AP}
\end{figure}

For offline analysis, our C++ software employs a derivative trigger to identify relevant signal events. Upon triggering, several basic parameters are computed:
\begin{itemize}
    \item Baseline level, baseline slope, and baseline Root Mean Square (RMS).
    \item Number of triggers within the acquisition window.
    \item Rise time, defined as the time difference between reaching 10\% and 90\% of the peak value.
    \item Decay time, defined as the time difference from 90\% to 30\% of the peak value.
\end{itemize}
These parameters are essential for selecting events to construct average pulse and noise templates to use the Optimum Filter technique  \cite{Gatti:1986cw} to improve the signal-to-noise ratio. To construct accurate average pulses, events are chosen based on the criteria of a single trigger in each pulse window, a flat pre-trigger trace, and a baseline RMS below a predetermined threshold. A typical signal from our GaAs cryogenic calorimeter is fully contained within a 70 ms window, as depicted in Fig.~\ref{fig:AP}.

To eliminate spurious events, waveforms are selected with single triggers and cuts on the rise and decay times that fall within a 3 $\sigma$around their average values, 1.2 ms and 10.8 ms for rise time and decay time, respectively.

\begin{figure}[h]
    \centering
    \includegraphics[width=0.5\textwidth]{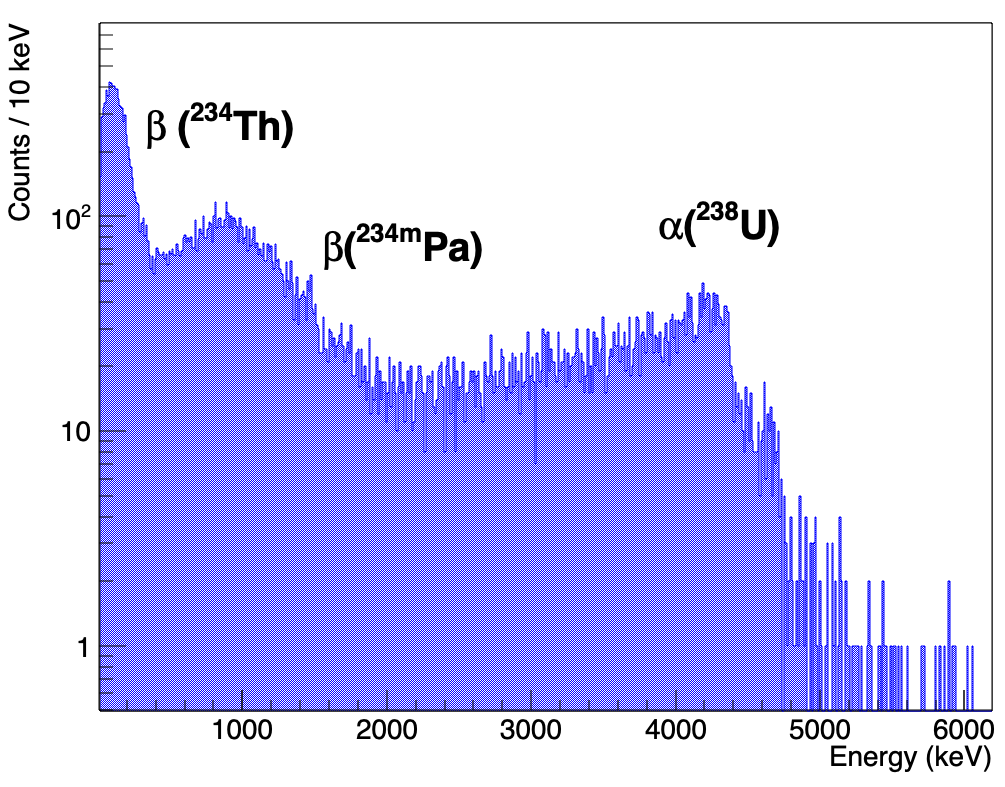}
    \caption{Energy spectrum of a 12 h calibration with the $^{238}$U Source from 0 to 6 MeV with PT on. This spectrum showcases key radioactive decay events originating from $^{238}$U. It highlights the quenched and degraded $\alpha$ events peaking at 4198 keV. At lower energy, there are the $\beta$ decays of $^{234}$Th with an endpoint at 274 keV and $^{234\text{(m)}}$Pm $\beta$ particles, with an endpoint energy at 2290 keV.}
    \label{fig:spectrum}
\end{figure}

The full energy spectrum of the events that passed the selection cut is presented in Fig.~\ref{fig:spectrum} calibrated with the  \textsuperscript{238}U $\alpha$ peak. 
The spectrum shows different populations that we could assess thanks to the choice of the calibration sources.  The choice of \textsuperscript{238}U source is strategic due to its emission of $\alpha$ particles at a prominent energy of 4198 keV, accompanied by $\beta$ particles from its daughter isotope, \textsuperscript{234}Th, with an endpoint energy of 2 MeV. These distinct energy signatures were deliberately chosen to facilitate a comprehensive calibration spectrum, offering a dual perspective on both high and relatively lower energy emissions within our detection range.
The choice of a smeared \textsuperscript{238}U source was instrumental in our approach, aiming to intentionally degrade the energy of $\alpha$ particles. This methodological decision resulted in a continuous energy spectrum, as opposed to discrete energy peaks, offering a broader range of energy deposition events for analysis. Such a spectrum is crucial for our future objectives, particularly in assessing our detector's particle identification capabilities under varying conditions. Fig.~\ref{fig:spectrum} distinctly showcases these features, providing visual confirmation of the anticipated $\alpha$ and $\beta$ emissions within our calibrated spectrum.

\begin{figure}[h]
    \centering
    \includegraphics[width=0.5\textwidth]{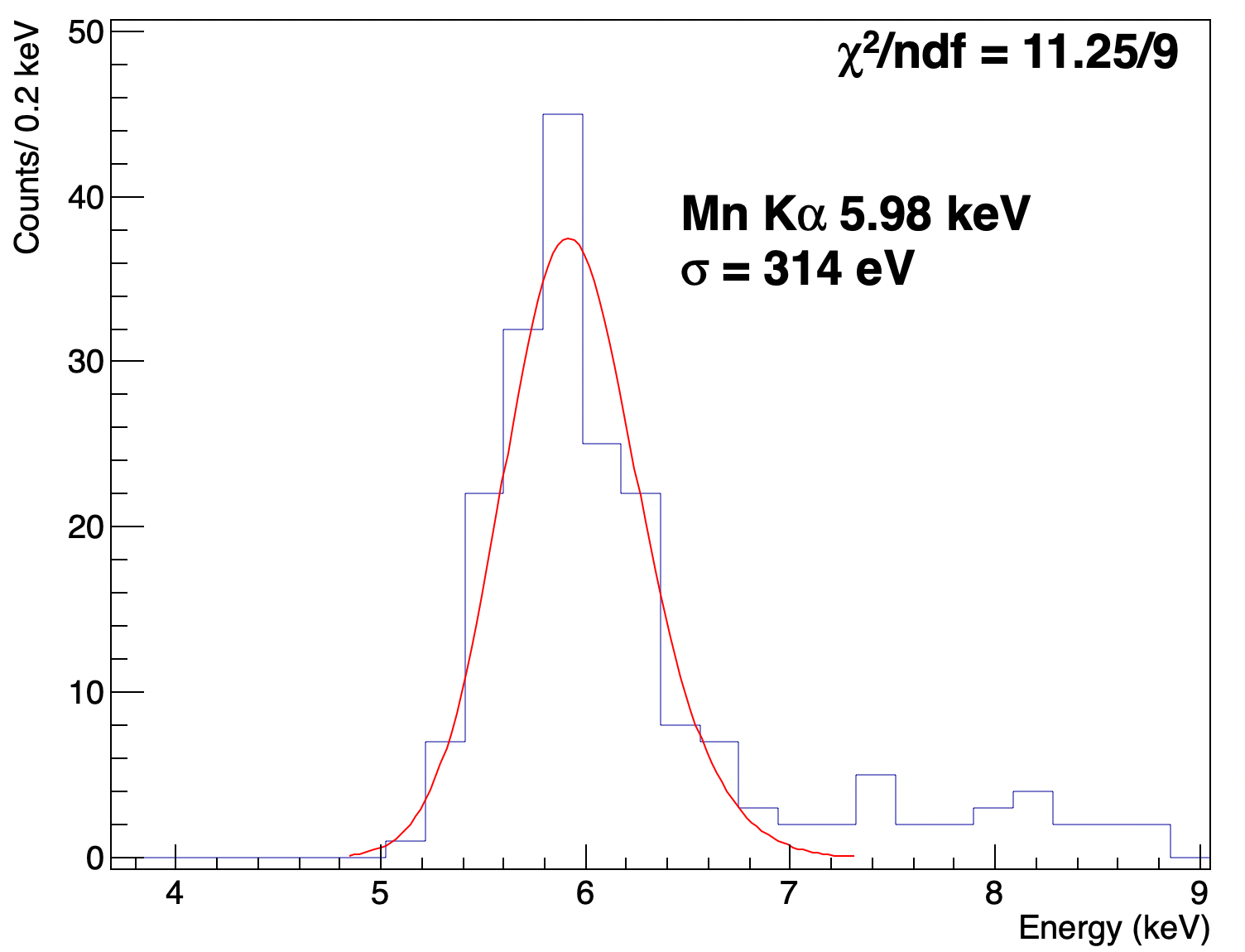}
    \caption{Low-energy X-ray spectrum obtained with the PT off, prominently featuring the 5.9 keV Mn K-$\alpha$ and 6.4 keV K-$\beta$ (indistinguishable). Notably, the energy resolution at the 5.9 keV Mn K-$\alpha$ peak is measured at 314 $\pm$ 22 eV extracted from the fit of the peak with two gaussian to account for Mn K-$\alpha$ and K-$\beta$.}
    \label{fig:spectrumfe}
\end{figure}

Taking advantage of the $^{55}$Fe source calibration, we could study the GaAs detector response at lower energies. The low energy spectrum of the GaAs is shown in Fig~\ref{fig:spectrumfe}. We obtained an energy resolution of 314 $\pm$ 22 eV ($\sigma$) at 5.9 keV and a baseline resolution of 283 $\pm$ 48 eV ($\sigma$) during a 2-hour-long run with the PT off. For the sake of comparison, the resolution at 5.9 keV during the 12-hour long (shown in Fig. \ref{fig:spectrum}) run with the PT on was 546 $\pm$ 21 eV ($\sigma$) and the baseline resolution was 542 $\pm$ 6 eV.
We summarize the main performances of the GaAs detector in Tab.~\ref{tab:perf}. These results shows that the main source of noise is from the PT. Indeed the experimental setup is not optimized for low threshold measurements.

\begin{table}[ht!]
   \centering

    \begin{tabular}{l c c}

\multicolumn{3}{c}{\textbf{Detector performance summary}} \\ 
\hline \hline
    Mass & 5.35 & g \\ \hline
    Density & 5.32 & g/cm$^3$ \\ \hline
    Diameter & 5.08 & cm \\ \hline
    Rise time 10-90  & 1.2 $\pm$ 0.1 & ms \\ \hline
    Decay time 90-30 & 10.8 $\pm$ 0.5 & ms \\ \hline
    NTD response & 450 & $\mu$V/MeV \\ \hline
    Baseline resolution (RMS) PT off & 283 $\pm$ 48 & eV \\ \hline
    Peak $\sigma$ at 5.9 keV PT off & 314 $\pm$ 22 & eV \\ \hline
    Baseline resolution (RMS) PT on & 542 $\pm$ 6 & eV \\ \hline
    Peak $\sigma$ at 5.9 keV PT on & 546 $\pm$ 21 & eV \\ \hline
   \end{tabular}
   \caption{Summary of the performance of GaAs detector operated as a low-temperature calorimeter.}
   \label{tab:perf}
\end{table}

Looking ahead, this calibration serves as a foundational step toward enhancing our experimental setup. The introduction of a cryogenic light detector in subsequent runs aiming to measure scintillation light, a pivotal component for particle identification. This advancement is expected to significantly refine our understanding of the detector's response to different types of radiation, thereby improving our ability to distinguish between particle interactions based on scintillation signatures.

In refining our analysis of the experimental setup, it is crucial to acknowledge that the detector was not operated in a low-radioactivity setup, since no effort was made towards selecting low-radioactivity materials. This operational choice, along with the high rate sources facing the detector for this first calibration measurement inevitably influenced the observed event rates, particularly contributing to an increased presence of events at lower energy i.e. below 1 MeV. The high event rate significantly impacted our ability to lower the detection threshold below the current 1 keV. The choice of two high-activity sources was aimed at performing a broad energy characterization of the detector response, i.e. from 1 keV to 4 MeV, to assess its performances over three orders of magnitude of signal amplitude.

\section{Conclusions and perspectives}
Our first test of this detector configuration at 15 mK was successful. The detector showed an energy resolution of $\simeq$ 546 eV and a baseline resolution  slightly below 542 eV, which sets a threshold  below 1.5 keV at 3$\sigma$ of the baseline noise.
These results are highly promising for the search for low-mass dark matter using GaAs crystals, demonstrating the potential of such detectors for this purpose. Despite the current limitations of the cryogenic facility and the use of an NTD sensor, the observed performance provides valuable insights into further improving the detector construction and setup.

GaAs is poised to become the inaugural cryogenic calorimeter featuring a triple readout capability of charge, phonons and scintillation light. As a future development we plan combining the GaAs crystal with a cryogenic light detector to measure scintillation light. Finally, we intend to install electrodes for charge collection, targeting the detection of individual electron-hole pairs generated during each interaction.

Moreover to further improve the threshold, we are considering Transition Edge Sensors as a thermal sensor. By coupling a TES with a metallic connection, specifically a bonding wire, directly to the GaAs, we anticipate significant improvements. Such improvements are expected to refine our measurement capabilities, pushing the boundaries of direct dark matter detection technology and opening up the possibility of exploring the currently inaccessible parameter space of the dark matter sector.

This work has been funded by the Italian University and Research Ministry thought the grant PRIN2022 - NextGenerationEU.
\section*{Acknowledgements}
We are particularly gratefully to Dr. J. Walker Beeman  for his invaluable support  in the  production of NTDs. We are also grateful to M. Guetti for the assistance in
the cryogenic operations abd the mechanical workshop of LNGS.

\section*{Availability of data and material}

Data will be made available on reasonable request to the corresponding author.

\bibliography{main.bib}
\bibliographystyle{spphys}

\end{document}